\documentclass[a4paper]{spie}  
\usepackage[utf8]{inputenc}

\usepackage{amsmath,amsfonts,amssymb}
\usepackage{astro_bib_macro_mam}

\newcommand{\beq}{\begin{equation}}
\newcommand{\eeq}{\end{equation}}
\newcommand{\beqna}{\begin{eqnarray}}
\newcommand{\eeqna}{\end{eqnarray}}

\newcommand{\hbx}{{\widehat{\boldsymbol{x}}}}

\newcommand{\hbc}{{\widehat{\boldsymbol{c}}}}

\newcommand{\bc}{{\boldsymbol{c}}}
\newcommand{\by}{{\boldsymbol{y}}}
\newcommand{\bx}{{\boldsymbol{x}}}
\newcommand{\be}{{\boldsymbol{e}}}
\newcommand{\beps}{{\boldsymbol{\epsilon}}}

\newcommand{\br}{{\boldsymbol{r}}}
\newcommand{\bs}{{\boldsymbol{s}}}

\newcommand{\bSigma}{{\boldsymbol{\Sigma}}}

\newcommand{\bI}{{\boldsymbol{I}}}

\newcommand{\bK}{{\boldsymbol{K}}}

\newcommand{\blambda}{\boldsymbol{\lambda}}

\usepackage{graphicx}
\usepackage[colorlinks=true, allcolors=blue]{hyperref}

\title{Statistical tests with multi-wavelength Kernel-phase analysis for the detection and characterization of planetary companions}

\author[a]{Mamadou N’Diaye}
\author[a]{David Mary}
\author[a]{Frantz Martinache}
\author[a]{Roxanne Ligi}
\author[a]{Nick Cvetojevic}
\author[a]{Peter Chingaipe}
\author[b]{Romain Laugier}
\affil[a]{Université Côte d'Azur, Observatoire de la Côte d'Azur, CNRS, Laboratoire Lagrange\\Bd de l'Observatoire, CS 34229, 06304 Nice cedex 4, France}
\affil[b]{Institute of Astronomy, KU Leuven, Celestijnenlaan 200D, 3001 Leuven, Belgium}

\authorinfo{Further information, send correspondence to \url{mamadou.ndiaye@oca.eu}}

\begin{document} 
\maketitle

\begin{abstract}
Kernel phase is a method to interpret stellar point source images by considering their formation as the analytical result of an interferometric process. Using Fourier formalism, this method allows for observing planetary companions around nearby stars at separations down to half a telescope resolution element, typically 20\,mas for a 8\,m class telescope in H band.
The Kernel-phase analysis has so far been mainly focused on working with a single monochromatic light image, recently providing theoretical contrast detection limits down to $10^{-4}$ at 200\,mas with JWST/NIRISS in the mid-infrared by using hypothesis testing theory.
In this communication, we propose to extend this approach to data cubes provided by integral field spectrographs (IFS) on ground-based telescopes with adaptive optics to enhance the detection of planetary companions and explore the spectral characterization of their atmosphere by making use of the Kernel-phase multi-spectral information. 
Using ground-based IFS data cube with a spectral resolution R=20, we explore different statistical tests based on kernel phases at three wavelengths to estimate the detection limits for  planetary companions. Our tests are first conducted with synthetic data before extending their use to real images from ground-based exoplanet imagers such as Subaru/SCExAO and VLT/SPHERE in the near future. Future applications to multi-wavelength data from space telescopes are also discussed for the observation of planetary companions with JWST. 
\end{abstract}


\section{INTRODUCTION}
\label{sec:intro}  

High-contrast observation is a burgeoning field in astrophysics to study the formation and evolution of planetary systems and survey the atmosphere of extrasolar planets. This field aims at imaging and spectrally analyzing faint companions around nearby stars, attempting to reach a contrast of several orders of magnitude (flux ratio $>10^3$) at angular separations smaller than one arc second (=1,000\,mas). Ground-based instruments rely on (i) adaptive optics (AO) for observed star images with a quality, i.e. Strehl ratio, smaller than 40\% and (ii) integral field spectrograph (IFS) for spatio-spectral data cube with medium resolution (R $\sim 4000$). Such means enable the observation of warm or massive Jupiters with $10^5$ contrast at separations wider than 300\,mas from their host stars \cite{Konopacky2013, Barman2015}.

Mid-resolution spectroscopy of exoplanets are relying on the spectral property differences between star and planet to ease the companion detection. Using cross-correlation methods between IFS data cube and molecule templates at the known planet location, these instruments have enabled the detection of absorption features such as methane (CH$_4$), water (H$_2$O), or carbon monoxide (CO) in the atmosphere of a few planets, providing constrains on their carbon-to-oxygen ratio. The additional location of the planetary companion with respect to its star and H$_2$O and CO snow lines gives insights on its formation and evolutionary mechanisms \cite{Oberg2011}.

These cross-correlation methods were recently proposed to be extended to all the pixels in the instrument field of view. Called molecular mapping, this strategy allows for simultaneously seeking planets and deriving spectral signatures \cite{Hoeijmakers2018}. Combining mid-resolution spectroscopy and high-contrast imaging offers additive contrast gains by spectrally and spatially filtering the light from an observed star to study its faint companions. Such a strategy has already been successfully probed on the archival data of VLT/SINFONI and KECK/OSIRIS to retrieve known imaged planets with unprecedented signal-to-noise ratio (up to 25)\cite{Hoeijmakers2018,Petit2018}.

While spectacular, these results in the near infrared (H \& K band, 1.65 and  2.2$\mu$m) were achieved with low instrumental transmission ($<$10\%), modest Strehl ratio ($<$40\%), and without starlight suppression device (i.e. coronagraph), therefore reducing sensitivity and preventing the observation of colder or lighter companions at 300\,mas. As spatial resolution is also related to the observing wavelength and telescope aperture size, detection is ultimately limited to companions with a separation larger than 40\,mas on 8\,m telescope in H-band. 

To overcome this limit, an alternative approach called Kernel phase looks at the image as if it were the result of an interferometric process \cite{Martinache2010,Martinache2013,Martinache2016,N'Diaye2018}. With the interferometric analysis of the observed star images with Fourier formalism, one can efficiently discriminate the stellar flux and probe companions down to 20\,mas. This approach has so far been employed on traditional well-corrected AO-corrected images (both narrow- and broad-band) to successfully retrieve relative astrometric and photometric parameters of companions. The method would benefit from the spectral coverage brought by an IFS with medium resolving power (R $\sim$ 4000) for two reasons: (1) the spectral resolution extends the capture range of the technique that becomes usable even when the correction is less than ideal \cite{Martinache2016b} and (2) one can build spectral differential kernels that make the observables even less susceptible to systematic errors that in practice impose contrast detection limits. Kernel and coronagraphy are orthogonal, observing planets at distinct separations.

Contrast detection limits have recently been derived for the kernel-phase analysis by using hypothesis testing theory \cite{Ceau2019}. In the context of JWST/NIRISS in full pupil mode, Ceau et al.\cite{Ceau2019} showed that contrasts of up to $10^{4}$ at separations of 200\,mas could ultimately be achieved with this approach, barring significant wavefront drift. These detection limits were given in the case of images at a single monochromatic light. Using kernel-phase analysis on multiple images at different position angles have recently been explored to push these detection limits further. First statistical tests are currently being designed on such a set of images (Ligi et al. in prep.). 

In this paper, we propose to extend the formalism of the kernel-phase analysis with statistical tests to multi-wavelength data. The formalism of the kernel phase approach is first recalled with a description of the derived observables. We then design three statistical detection tests in the presence of  multi-channel data to determine the probability of detection for a planetary companion. We determine the boundaries of the tests in the case of 3 spectral channels and compare results with the case of a single channel. 

\section{PLANETARY COMPANION DETECTION WITH KERNEL PHASE}

We here recall the principle of the Kernel-phase analysis. Further details can be found in the literature\cite{Martinache2010,Martinache2020}.

We assume the observation of an unresolved star with a telescope and small residual aberrations, i.e. a phase $\varphi \ll 1$. This configuration can be represented by a space mission or a ground-based observatory using adaptive optics with good correction of the effects of the atmosphere on the image of an observed star. The electric field $E_P$ at the telescope pupil $P$ can then be written as

\begin{equation}
    E_P = P \exp (i\varphi)\,,
\end{equation}
in which $\varphi$ denotes the phase residuals. In the following image plane, the star electric field is expressed as

\begin{equation}
    E_{im} = \mathcal{FT}(P \exp (i\varphi))\,,
\end{equation}
in which $\mathcal{FT}(f)$ represents the Fourier transform of $f$. The intensity is then simply given by the following expression
\begin{equation}
    I_{im} = | E_{im} |^2\,.
\end{equation}
Assuming an unresolved point source, this term represents the point-spread function (PSF) of the instrument and its environment. The optical transfer function (OTF) is obtained with a Fourier transform of the previous expression. By taking the argument of the OTF, we obtain the phase $\Phi$ in the Fourier plane. The phases in this Fourier space and the pupil plane are linearly related in the regime of small aberrations ($\varphi \ll 1$\,rad). Using pupil discretization, this relation at the spectral channel (i) can be expressed as
\begin{equation}
    \mathbf{\Phi}^{(i)} = \mathbf{R}^{-1}\mathbf{A}\mathbf{\varphi}^{(i)} + \mathbf{\Phi}_0^{(i)}\,. 
    \label{eq:kerneleq}
\end{equation}
where $\mathbf{A}$ represents the baseline mapping matrix between both phases, $\mathbf{R}$ denotes the diagonal matrix representing the redundancy of the subaperture pairs that contirbute the Fourier phase for that baseline, and finally $\mathbf{\Phi}_0$ represents the signature In the presence of a planetary companion. At an angular separation ($\alpha,\beta$), the companion's phase signature  vector in channel $i$ is of the form
 \begin{equation}
\boldsymbol{\Phi_0}{^{(i)}}=\angle{\left (1+\bc_i \exp{\left (-\displaystyle{\frac{2i\pi}{\lambda_i}}(\alpha\boldsymbol{u}+\beta\boldsymbol{v})\right )}\right )}.
\label{phi0i}
\end{equation}
in which $c_i$ defines the flux ratio of the planetary companion with respect to the star at the channel (i), ($\mathbf{u}$,$\mathbf{v}$) represent the coordinate vectors in the Fourier space, and $\angle$ denotes the argument operator.

In the context of a faint planetary companion close to a bright star, we need to extract the signature of the planet from the star signal represented by the product $\mathbf{R}^{-1}\mathbf{A}\mathbf{\varphi}^{(i)}$. We can look for the kernel operator $\mathbf{K}$ such that
\begin{equation}
    \mathbf{K}\mathbf{R}^{-1}\mathbf{A} = \mathbf{0}\,.
\end{equation}
The so-called Kernel matrix $\mathbf{K}$ cancels out the great majority of the aberrations that alters the phase present in the Fourier space, while keeping the information about the planetary companion. By applying this matrix to the equation (\ref{eq:kerneleq}), we obtain
\begin{equation}
    \mathbf{K}\mathbf{\Phi}^{(i)} = \mathbf{K}\mathbf{\Phi}_0 ^{(i)}\,. 
\end{equation}
In this case, we obtain the so-called Kernel-phase observables with $\mathbf{k}^{(i)}=\mathbf{K}\mathbf{\Phi}^{(i)}$ and $\mathbf{k}^{(i)}_0=\mathbf{K}\mathbf{\Phi}^{(i)}_0$. The vector has a size of $p$. Assuming some noise correlation between the different components of the kernel phase, we can set a residual noise $\mathbf{n}^{(i)}$ as a multi-variate Gaussian noise distribution with a matrix average $\mathbf{0}$ and a covariance matrix $\mathbf{\Sigma}^{(i)}$. The expression can be written as
\begin{equation}
    \mathbf{k}^{(i)} = \mathbf{k}^{(i)}_0 + \mathbf{n}^{(i)}\,.
\end{equation}

If we have enough data, it will be possible to build this covariance matrix and derive the corresponding whitening matrix. In this case, it is convenient to work with the whitened version of the observable $\mathbf{y}$ that can be obtained as follows
\begin{equation}
    \mathbf{y}^{(i)} = \mathbf{\Sigma^{(i)}}^{-\frac{1}{2}} \mathbf{k}^{(i)}\,.
\end{equation}

This leads to the following statistical model in the absence (under the null hypothesis $\mathcal{H}_0$) or in the presence of a companion ($\mathcal{H}_1$).
\begin{equation}
\begin{cases}
\mathcal{H}_0: \by{^{(i)}}= \br{^{(i)}} + \beps^{(i)},\\
\mathcal{H}_1: \by{^{(i)}}=\bx{^{(i)}}+  \br{^{(i)}} + \beps{^{(i)}}, 
\end{cases}
\beps{^{(i)}}\sim \mathcal{N}(\boldsymbol{0},\bI), \quad i=1,\cdots, q,
\label{finalmodel0}
\end{equation}
in which
\begin{equation}
\bx{^{(i)}}:={\boldsymbol{\Sigma}^{(i)}}^{-\frac{1}{2}} \boldsymbol{k}_0^{(i)}  =  {\boldsymbol{\Sigma}^{(i)}}^{-\frac{1}{2}}\boldsymbol{K}\angle{\left (1+\bc_i \exp \left (-\textrm{i}\frac{2\pi}{\blambda_i}(\alpha\boldsymbol{u}+\beta\boldsymbol{v})\right )\right )}\,,
\label{comp}
\end{equation}
 is the companion's  whitened kernel phase signature, $\br{^{(i)}}= {{\boldsymbol{\Sigma}}^{(i)}}^{-\frac{1}{2}}\be'{^{(i)}}$ is the residual whitened model error,  and $\bI$ is the $p \times p$ Identity matrix.

If the residual errors can be estimated with measurements from a calibration star, they can be further removed, leading to the idealistic calibration error-free model:
\begin{equation}
\begin{cases}
\mathcal{H}_0: \boldsymbol{y}{^{(i)}}=\beps{^{(i)}},\\
\mathcal{H}_1: \boldsymbol{y}{^{(i)}}=\boldsymbol{x}{^{(i)}}+\beps{^{(i)}}.
\end{cases}
\beps{^{(i)}}\sim \mathcal{N}(\boldsymbol{0},\bI), \quad i=1,\cdots, q\,.
\label{finalmodel}
\end{equation}

For subsequent uses, we define the vector
concatenating the companion kernel signatures at all wavelengths by $\bx$ : 
\begin{equation}
\bx := [\bx{^{(1)}}^\top \cdots \bx{^{(q)}}^\top ]^\top. \quad
\end{equation}
Similarly, we will denote by $\by$ and $\beps$ the vectors  concatenating the corresponding components in all channels.

\section{STATISTICAL HYPOTHESES AND KERNEL-PHASE DETECTION LIMITS}

We here analyse the detection limits for planetary companions with the kernel phase analysis by investigating three different tests. While Ceau et al.\cite{Ceau2019} presented detection tests for data from a single channel, our work extends the approach to multi-channel data. 

\subsection{Neyman-Pearson (likelihood ratio) test}
For the hypothesis problems described above, the most powerful test is the so-called likelihood ratio or Neyman-Pearson test\cite{Neyman1933}, defined as
\begin{equation}
\frac{\ell(\boldsymbol{x};\boldsymbol{y})}{\ell(\boldsymbol{0};\boldsymbol{y})}\underset{\mathcal{H}_0}{\overset{\mathcal{H}_1}{\gtrless}} \eta\ ,
\label{eq:LR}
\end{equation}
in which $\ell(\boldsymbol{x};\boldsymbol{y})$ is the likelihood and $\eta$ is a threshold allowing us to tune the false alarm rate. Assuming that the data from each channel are independent, the global likelihood is the product of the likelihood of data vector at each wavelength:
\begin{equation}
\ell(\boldsymbol{x};\boldsymbol{y})=\displaystyle{\prod_{i=1}^{q}} \ell(\boldsymbol{x}^{(i)};\boldsymbol{y}^{(i)})\,. 
\label{prod}
\end{equation}
Following Ceau et al.\cite{Ceau2019}, the test leads to
\begin{equation}
T_{\mathrm{NP}}(\boldsymbol{y},\boldsymbol{x})=\boldsymbol{y}^T\boldsymbol{x} \underset{\mathcal{H}_0}{\overset{\mathcal{H}_1}{\gtrless}} \xi,
\label{eq:NP}
\end{equation}
where $\xi$ is a threshold allowing us to tune the false alarm rate.
The Neyman-Pearson test uses the scalar product of the data with the true vector in the presence or in the absence of a planetary companion to compare it with a given threshold. By setting the latter to a given value, we can determine whether there is a companion or not for a given false probability alarm.

\subsection{Energy detector test}
The Generalized Likelihood Ratio (GLR) test arises from a plug-in strategy, where the unknown parameters are replaced by their estimates and injected in the likelihood ratio. In the particular case in which the signature is fully unconstrained, we obtained the estimated $\hbx = \by$, and the GLR leads to the energy detector test: 

\begin{equation}
T_{\mathrm{E}}(\boldsymbol{y})=||\boldsymbol{y}||^2\underset{\mathcal{H}_0}{\overset{\mathcal{H}_1}{\gtrless}} \xi\,.
\label{eq:ED2}
\end{equation}

\subsection{Binary detector test}
In a more focused approach, the maximum likelihood estimates (MLE) may be searched in the space $\mathcal{X}$
of admissible signatures for a companion. In this case,
 there are $2+q$ free parameters: $\alpha$, $\beta$ and the entries of contrast vector $\bc \in \mathbb{R}^{q}$. Using \eqref{comp} and \eqref{prod}, the MLE becomes
\begin{equation}
\begin{split}
\hbx:&=\arg\ \underset{\alpha,\beta, \bc}{\mathrm{  sup}\ } \ell(\alpha,\beta,\bc;\boldsymbol{y})\\
&=\arg\ \underset{\alpha, \beta, \bc}{\mathrm{  sup}\ }
\displaystyle{\prod_{i=1}^{q}}
{e^{\displaystyle{-\frac{1}{2}||\boldsymbol{y^{(i)}}-\boldsymbol{\Sigma^{(i)}}^{-\frac{1}{2}}\boldsymbol{K}\angle{(1+\bc_i e^{\displaystyle{-\textrm{i}\frac{2\pi}{\blambda_i}(\alpha\boldsymbol{u}+\beta\boldsymbol{v})}})}||^2}}}.\\
\end{split}
\label{eq:MLE2}
\end{equation}
In this case, finding the MLE requires numerical methods because it cannot be obtained analytically. 

In the case of small contrasts, the parameters space 
can however be reduced from $q+2$ to $2$ since the $q$ MLE contrast estimates can be obtained analytically. Indeed, in this case, Ceau shows in his thesis'\,appendix B that
the companion's signature at a particular wavelength $\blambda_i$ 
can be approximated to the first order as 
\begin{equation}
\bx^{(i)} \approx  \bc_i\bs^{(i)} \quad \textrm{with \quad} \bs^{(i)} = {\bSigma^{(i)}}^{-\frac{1}{2}}\bK \sin(-\frac{2 \pi}{\blambda_i}(\alpha \boldsymbol{u} + \beta \boldsymbol{v})).
\end{equation}

For a given couple $(\alpha,\beta)$, the global minimization problem amounts, for
small contrasts, 
finding the contrasts' MLE in each channel:
\begin{eqnarray}
\hbc_i(\alpha,\beta) &=& \arg\ \underset{\bc_i>0}{\mathrm{  sup}\ } \ell(\alpha,\beta,\bc_i;\boldsymbol{y})\\
& = & \arg\ \underset{ \bc_i>0}{\mathrm{  sup}\ } 
{e^{\displaystyle{-\frac{1}{2}||\boldsymbol{y^{(i)}}-   \bc_i\bs^{(i) }  } \|^2}},\\
 & = & \arg\ \underset{ \bc_i>0}{\mathrm{  inf}\ } \| \by^{(i)} - \bc_i \bs^{(i)} \|^2,
\end{eqnarray}
in which we have made explicit the constraint that the contrast is positive (but  the companion's spectrum is otherwise  unconstrained). This  leads to
\beq
\hbc_i(\alpha,\beta) = \max\left(\frac{{\bs^{(i)}}^\top \by^{(i)}}{{\bs^{(i)}}^\top {\bs^{(i)}}},0\right),
\label{estimci}
\eeq
in which we have emphasized the dependence of the optimal contrast in $(\alpha,\beta)$. For small contrasts, the likelihood involved in the numerator of GLR test hence writes as
\begin{eqnarray}
 \ell(\hbx;\by) & = & 
  \underset{\alpha,\beta}{\mathrm{  sup}}\   \displaystyle{\prod_{i=1}^{q}} 
  {e^{\displaystyle{-\frac{1}{2}||\boldsymbol{y^{(i)}}-   {\hbc_i}(\alpha,\beta)\bs^{(i) }  } \|^2}}.\\
\end{eqnarray}

Injecting this in the GLR test statistic and applying the positivity constraint at each channel yields
 \begin{eqnarray}
 T_B 
& =& \underset{\alpha,\beta}{\mathrm{  sup}}\ \displaystyle{\sum_{i=1}^{q}} 
\max\left(\frac{{\bs^{(i)}}^\top \by^{(i)}}{\sqrt{{\bs^{(i)}}^\top {\bs^{(i)}}}},0\right).
\label{tb}
 \end{eqnarray}

\section{MULTI-WAVELENGTH KERNEL-PHASE ANALYSIS}

\subsection{Assumptions}
For our preliminary simulations, we adopted the following set of parameters. For the astrophysical scene, we assume a F0-type star with a magnitude $m_{H}=5$ in H-band and a T-type planerary companion with a magnitude difference $\Delta m_{H}=8$, corresponding to a contrast ratio of about $6.3\times 10^{-4}$ with its parent star. The angular distance between the star and planet is set to 100\,mas. 

The simulated images are produced with the Subaru telescope of diameter $D$=7.92\,m and its SCExAO high-contrast instrument\cite{Jovanovic2016,Lozi2018}. The overall telescope and instrument transmission is assumed to be about 10\%. In terms of noise source, we include the presence of atmospheric residuals of about 100\,nm root mean square (RMS) after correction by adaptive optics, photon noise and a 10\,e$^{-}$ RMS readout noise. For this batch of simulations, no sky transmission and emission is considered here. We work a set of 1000 realizations of short-exposure images. In further work, we will consider long-exposure time images to gain consistency with real IFS data cube. 

\begin{figure} [!ht]
   \begin{center}
   \begin{tabular}{c} 
   \includegraphics[height=6cm]{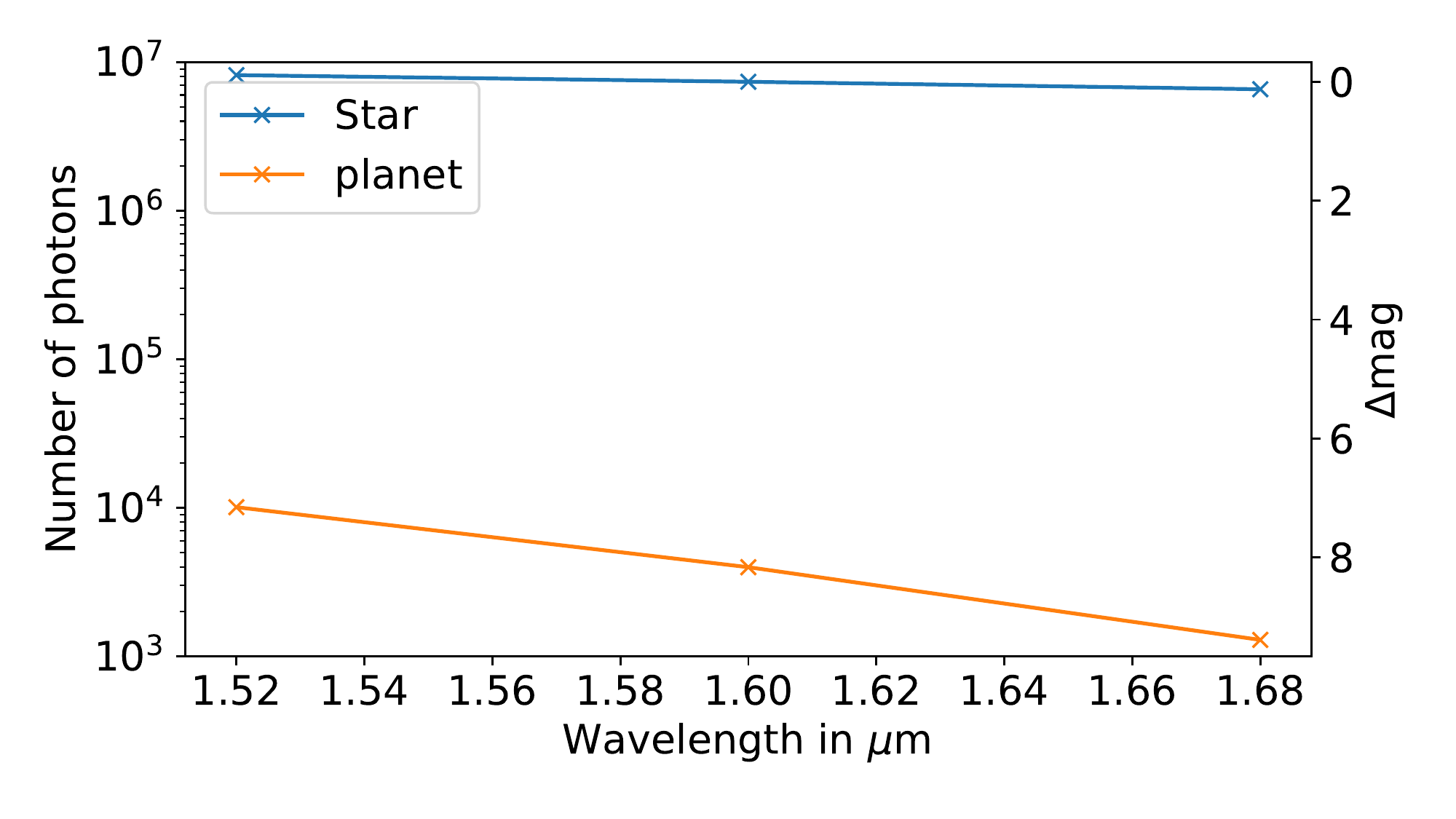}\\
   \includegraphics[height=6cm]{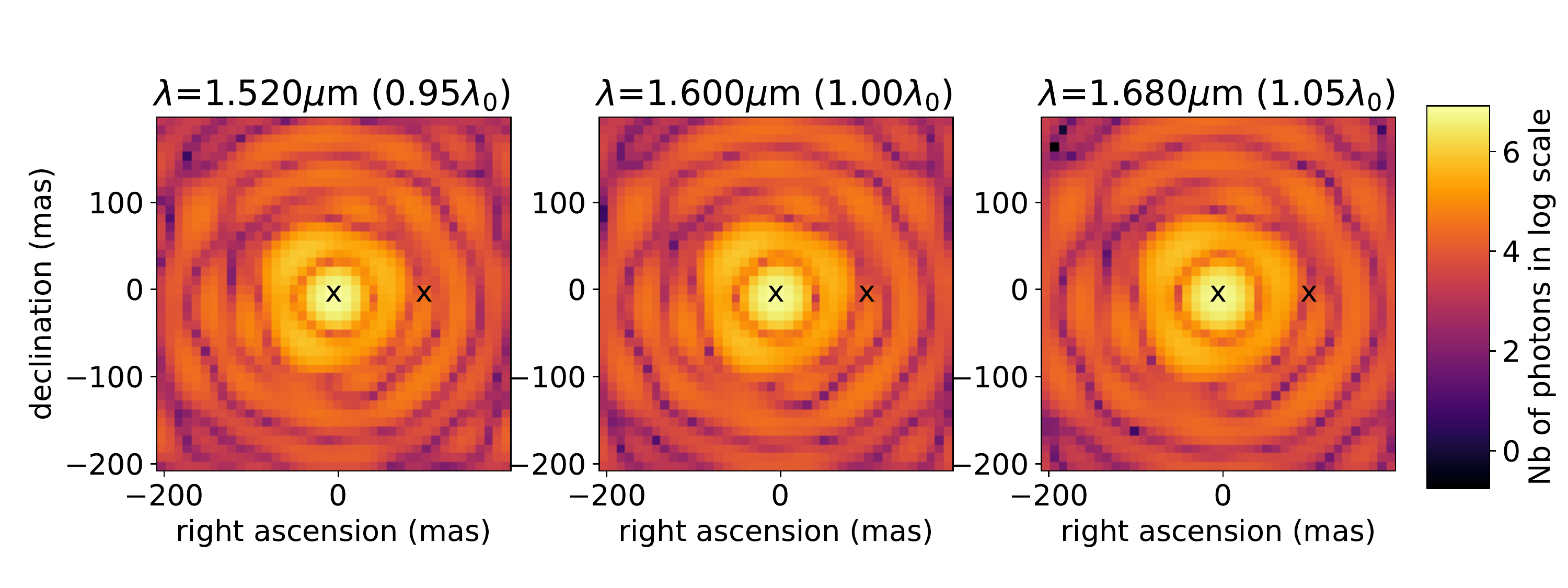}
   \end{tabular}
   \end{center}
   \caption[example] 
   { \label{fig:SED_star_planet} 
\textbf{Top}: Spectral energy distribution of the F0-type star and the T-type planetary companion with a spectral resolution R=20 for three spectral channels in H-band. \textbf{Bottom}: example of simulated images of the astrophysical scene at the three spectral channels in the presence of AO residuals, photon noise, and readout noise. The crosses denote the position of the star at the center and the position of the companion at 100\,mas with a contrast of $6.2\times 10^{-4}$.}
   \end{figure}

In terms of spatial sampling, the images have a size of 40$\times$40 spaxels with a plate scale of 10.42\,mas/spaxel, leading to 4 spaxels per resolution $\lambda_0/D$ in which $\lambda_0$ denotes the central wavelength of the considered bandwidth, here $\lambda_0=1.6\mu$m. The spectral resolution $R$ is 20, which is equivalent to the CHARIS IFS module of SCExAO at the low-resolution mode. In terms of spectral sampling, the spectral coverage $\Delta\lambda/\lambda$ is set to 10\%, leading to three monochromatic light channels for our simulations: 1.52, 1.60 and 1.68\,$\mu$m within the spectral band. Figure \ref{fig:SED_star_planet} represents the spectral energy distribution of the star and planet for our simulation in the top plot and a set of images at these three spectral channel. 

Following aperture prescriptions to minimize calibration errors with the kernel-phase analysis\cite{Martinache2020}, the Subaru telescope is modeled with 320 sub-pupils, leading to a number of 622 baselines and a number of 303 kernel-phase per spectral channel. To develop our statistical model, we first compute the covariance matrix from the kernel-phases achieved with the 1000 realizations of images at each wavelength in the presence and in the absence of our planetary companion. We then compute the whitening matrix to whitened the kernel phases for our observable $\mathbf{y}^{(i)}$ for each channel and for both hypothesis. 
To develop our statistical model, we finally perform Monte-Carlo simulations with 30,000 realizations of whitened kernel phases, using a residual error noise $\textbf{epsilon}^{(i)}$ following a Gaussian distribution $\mathcal{N}(\boldsymbol{0},\bI$). 
The multi-channel whitened kernel-phase are then concatenated to probe the different statistical tests.

\subsection{Multi-channel results}
\begin{figure} [!ht]
   \begin{center}
   \begin{tabular}{ccc} 
   \includegraphics[height=8cm]{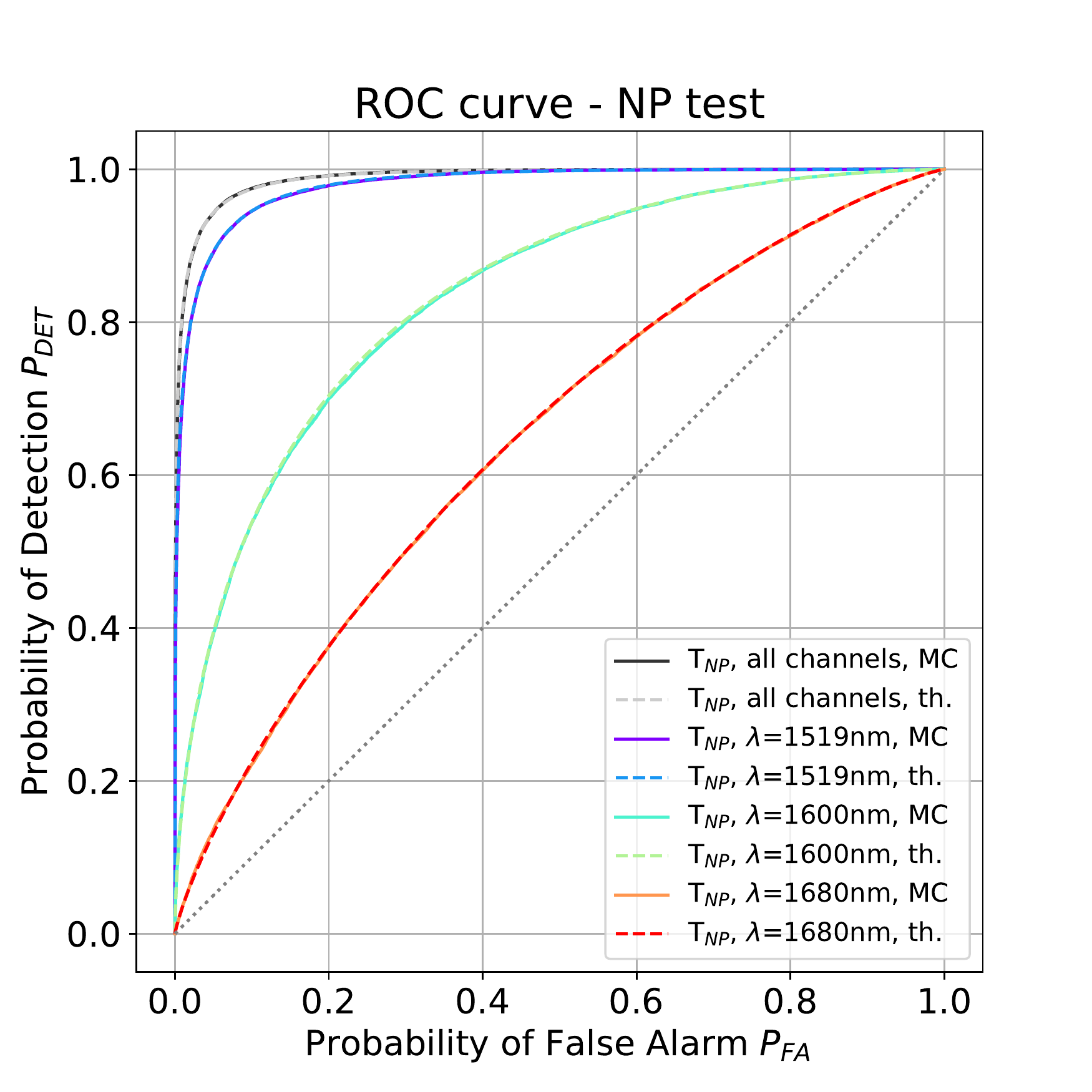}
    \includegraphics[height=8cm]{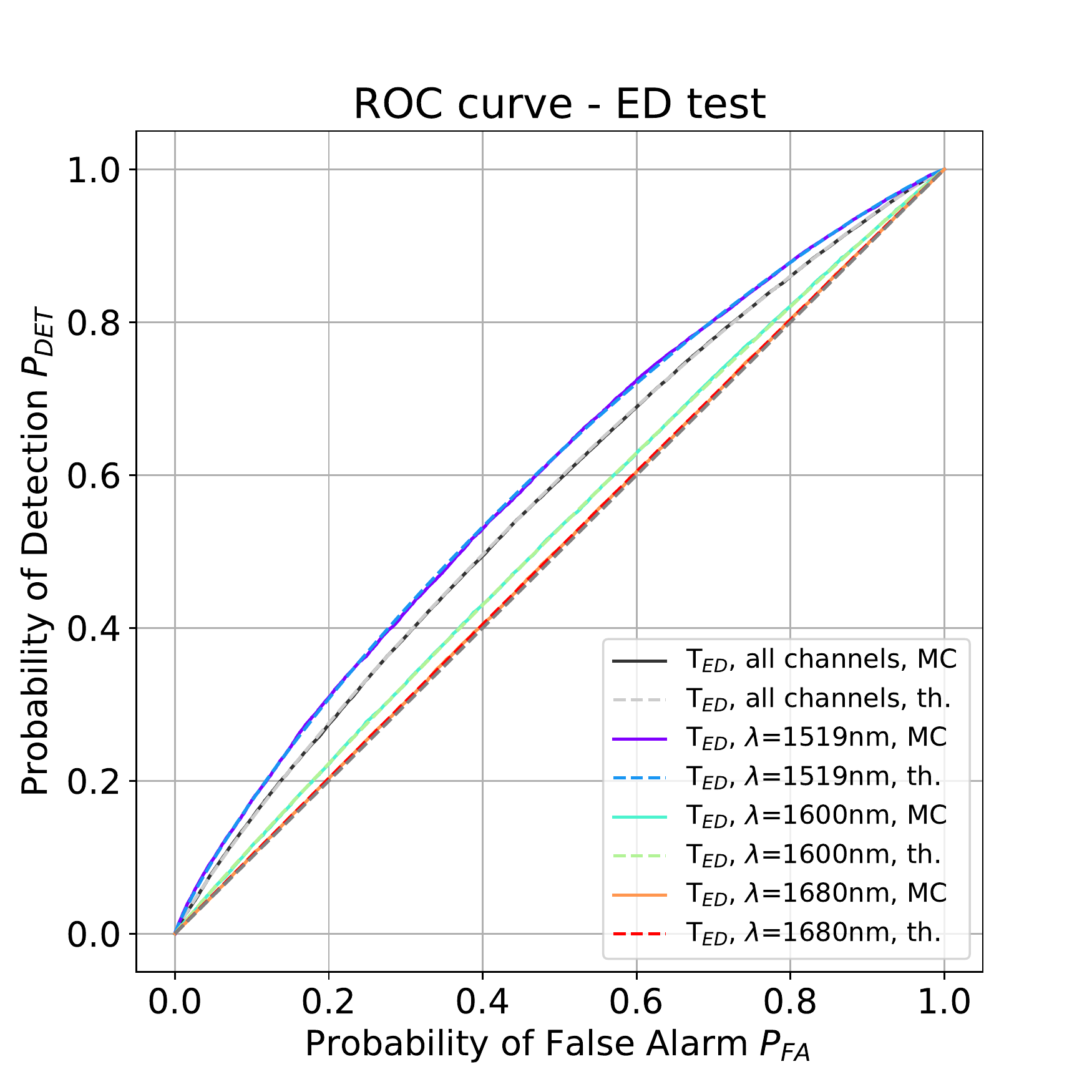}\\
    \includegraphics[height=8cm]{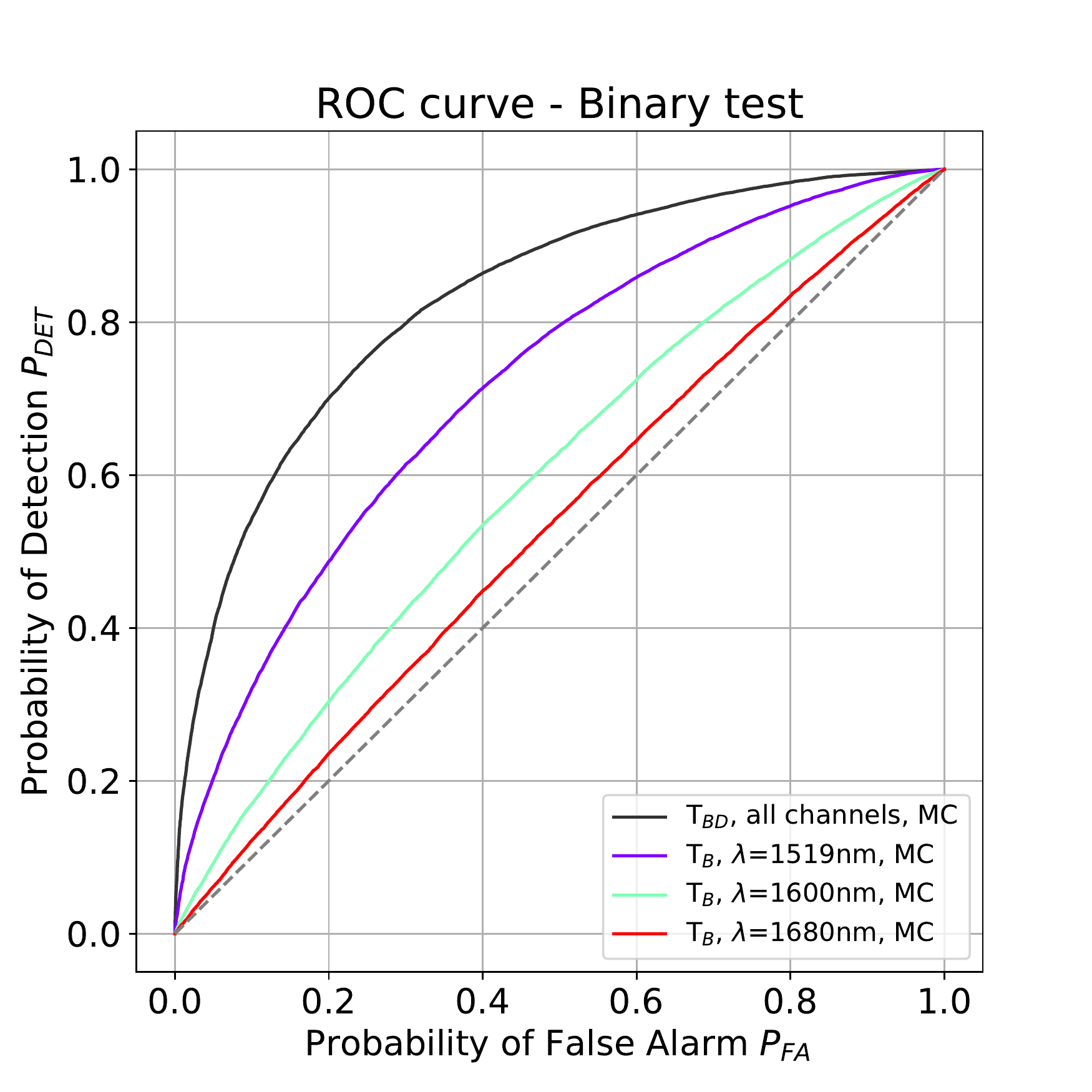}
   \end{tabular}
   \end{center}
   \caption[example] 
   { \label{fig:ROCcurves_per_test} 
ROC curves for the Neyman-Pearson ($T_{NP}$), the energy detector ($T_{ED}$) and the binary detector ($T_{B}$) tests on the top left, top right and bottom plots in the case of a planet at 100\,mas from its host star and an averaged magnitude $\Delta m_{H}=8$ in H-band. The solid line curves represent the experimental data derived from our Monte-Carlo simulations of whitened kernel-phase observables in multi-channel (black) and single-channel (blue, green, and red) cases at different wavelengths. For the $T_{NP}$ and $T_{ED}$, the dashed lines represent the curves from derived from theory, showing excellent agreement with the experimental data. The diagonal dotted line represents the random classifier.}
   \end{figure}

Figure \ref{fig:ROCcurves_per_test} shows the results of our statistical curves in the form of Receiver Operating Characteristic (ROC) curves for the Neyman-Pearson ($T_{NP}$), Energy detector ($T_{ED}$), and Binary detector ($T_B$) tests, at respectively the top left, top right, and bottom plots. For each plot, we represent the results for the test considering all the three channels and the channels individually. In the case of the $T_{NP}$ and $T_{ED}$, we also add the theoretical curves that can be derived from an analysis that is similar to the one showed in Ceau et al.\cite{Ceau2019}. In all cases, the experimental data show good agreement with the theoretical curves.

For the Neyman-Pearson test (top left plot), it is interesting to note the improvement of the test with the multi-channel approach with respect to the single wavelength approach, showing the interest in combining data from several spectral channels to enhance detection limits. 
We can also notice a difference between the ROC curves from the three individual channels, the ROC curves moving towards the perfect classifier as the wavelength shifts toward the blue end of the spectrum. This difference most likely comes from the evolution of the contrast between the star and the planet as a function of the wavelength and showed in Figure \ref{fig:SED_star_planet} top plot. In our example, the contrast is more favorable towards the shortest wavelength, therefore maximizing the chance to detect the companion with the whitened kernel phase observables as the contrast increases.

For the Energy detector test (top right plot), the observed behavior is different from the previous test. The ROC curve is more favorable at the shortest wavelength than the multi-wavelength channel case. If a spectral feature is much stronger at a given wavelength, the mono-channel test may prove more interesting. For this unconstrained GLR case, we envision to perform further tests to determine at which strength of a spectral feature it may become more interesting to look at multiple instead of single channel test. 

Finally for the binary test (bottom plot), the observed behavior is similar to the one observed with the Neyman-Pearson test. With the binary detector test, we can clearly see an improvement by summing up the information of several channels compared with single channel ROC curves.

\begin{figure} [!ht]
   \begin{center}
   \begin{tabular}{c} 
   \includegraphics[height=8cm]{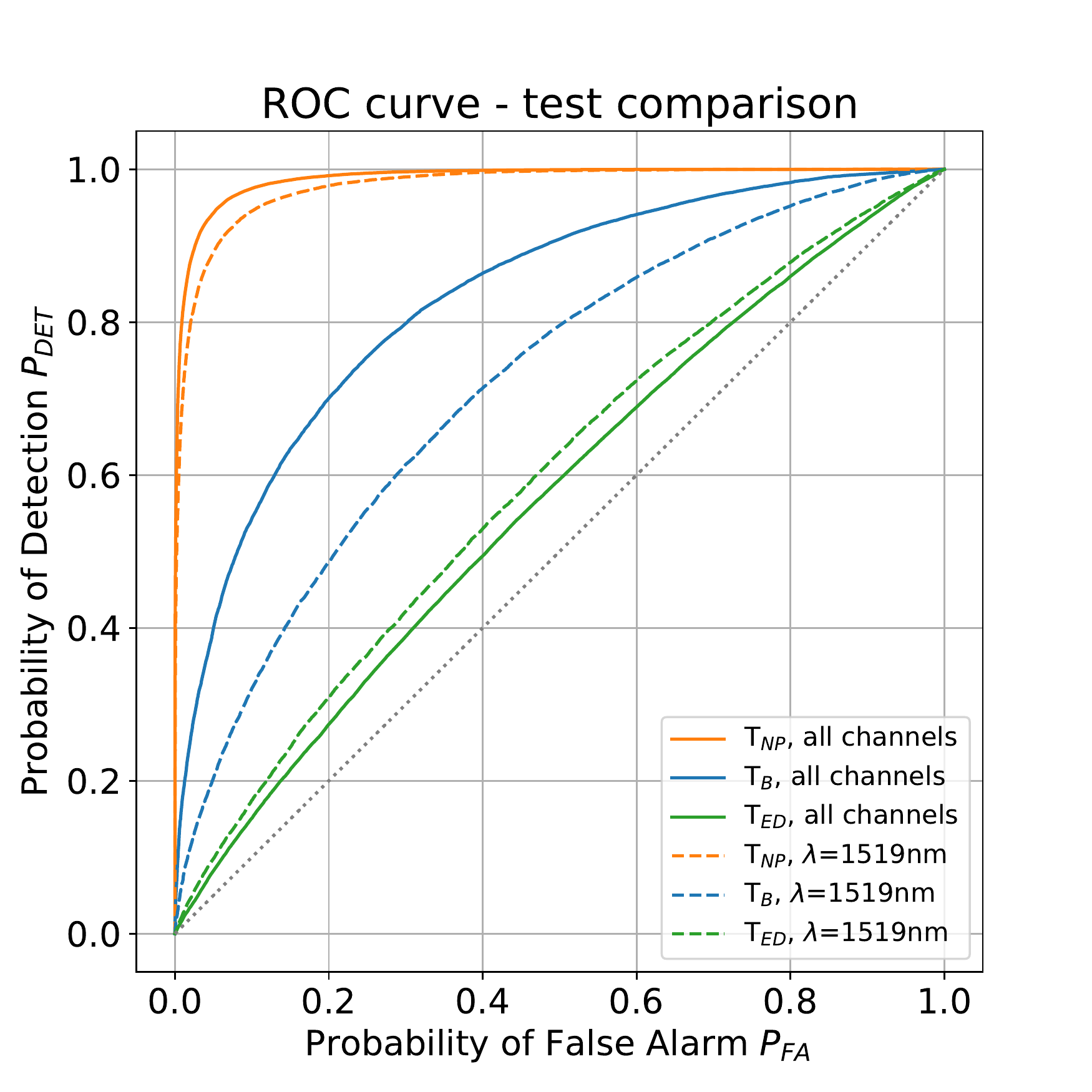}
   \end{tabular}
   \end{center}
   \caption[example] 
   { \label{fig:ROCcurves_monopoly} 
ROC curves for the Neyman-Pearson ($T_{NP}$, orange), the energy detector ($T_{ED}$, green) and the binary detector ($T_{B}$, blue) tests in the case of a planet at 100\,mas from its host star and an averaged magnitude $\Delta m_{H}=8$ in H-band. The curves represent the experimental data derived from our Monte-Carlo simulations of whitened kernel-phase observables in multi-channel (solid line) and single-channel (dashed-line) cases at 1519\,nm. The diagonal dotted line represents the random classifier.}
   \end{figure}

We then look at a comparison of the results between the different tests, see Figure \ref{fig:ROCcurves_monopoly}. For these plots, we show the ROC curves for the multi-channel case and the shortest wavelength case which is the most favorable in our simulation data set. As expected from theory, the Neyman-Pearson test is the most powerful both in single and multi-channel cases. It is interesting to note the overall improvement observed with the binary detector test which remains more powerful than the energy detector test. Except for the energy detector, an overall improvement is observed with he multi-channel statistical tests compared with the monochromatic cases.

In further works, it will be interesting to compare these statistical tests in single and multiple channels by exploring the parameter space for the star-planet couple for different contrast, separation, star magnitude, noise regimes and different spectral sampling.

\section{CONCLUSIONS AND PROSPECTS}
In this paper, we have developed a formal multi-wavelength detection approach to analytically derive the upper detection limit for the kernel phase with the Neaman-Pearson test. We show the control of the false alarm rate for all the considered tests (Neaman-Pearson, energy detector, and binary detector). Our first results have showed that a gain is already observable with three channels over one channel. A larger gain is expected with an increased number of channels. 

In the contribution, we limited our work to three tests. We are currently designing even more pewerful tests to leverage some specific spectral features, which could prove interesting for both detection and spectral analysis of the atmosphere of a planetary companion. Our next step will then be to perform a validation of our tests with on-sky data from with ground-based exoplanet imagers on 8\,m class telescopes such as Subaru/SCExAO\cite{Lozi2018}, VLT/SPHERE\cite{Beuzit2019} or Gemini Planet Imager\cite{Macintosh2014}. We also expect to extend these studies on the on-going JWST cycle 1 program on the survey of 20 Y dwarfs \cite{Albert2021}. 


\acknowledgments 
This work was supported by the Action Spécifique Haute Résolution Angulaire (ASHRA) of CNRS/INSU co-funded by CNES.
 
\bibliography{2022_mndiaye_biblio} 
\bibliographystyle{spiebib} 

\end{document}